# CdTe$_{0.25}$Se$_{0.75}$ Quantum Dots as Efficient Room Temperature Single Photon Source for Quantum Technology

*Kush Kaushik, Jiban Mondal, Ritesh Kumar Bag, Shagun Sharma, Abdul Salam, Chayan Kanti Nandi[*]*

School of Chemical Sciences, Indian Institute of Technology Mandi, Mandi, H. P. 175075, India

[*]Corresponding author email: chayan@iitmandi.ac.in

## Abstract

Room temperature single photon sources (SPS) are crucial for developing the next generation quantum technologies. Quantum dots (QDs), recently, have been reported as promising materials as SPS at room temperature. By optimizing the single particle optical properties of a series of water-soluble, CdTe$_x$Se$_{1-x}$, here we provide an efficient SPS with increased single photon purity. The data revealed that second order photon correlation, g$^2$(0) value decreases substantially from 0.21 in CdTe to 0.02 in CdTe$_{0.25}$Se$_{0.75}$ QDs. They also exhibited deterministic emissions with an increase in ON time exceeding 95% of the total time. This was accompanied by an increased photon count rate, substantially reduced blinking events, and extended single particle ON-time. The increased single photon emission in CdTe$_x$Se$_{1-x}$ is attributed to very fast electron trapping to dense trap states, which suppresses the multiexciton recombination.

**Introduction**

Single photon sources (SPS) are obvious requirement for the quantum technologies which have been revolutionizing in recent years due to their superior applications in quantum technology, such as quantum computing[1], quantum communications[2,3], quantum sensing[4] and metrology[5,6] and quantum information processing[7,8]. Quantum computing is often achieved with the help of two-level quantum systems like superconducting circuits[9,10], trapped ions[11], electron spin in an atom or quantum dot[12,13] etc. Unfortunately, most of these systems lose quantum coherence due to interactions between the quantum systems and environmental factors like thermal fluctuations, electromagnetic fields or other particles[14,15]. A variety of materials like NV center in diamond[16], organic molecules[17], carbon nanotubes[18], 2D materials (hBN, $MoSe_2$, $WS_2$ etc.)[19], QDs[16,20] have been explored for exploiting as SPS. However, all these materials are often operated at ultra-low temperatures (millikelvin to a few Kelvin) to suppress these effects and to obtain the single photon emission[21]. Maintaining such sophisticated facilities produces several challenges in terms of scalability and usability, as well as costs incurred. Therefore, the development of room temperature single photon source is of utmost importance to circumvent these issues.

An efficient room temperature SPS still remains a great challenge. Out of several important factors, efficiency of SPS is typically measured using minimizing the photon anticorrelation effect of single photons. A single photon cannot be detected simultaneously at two different locations and hence, it must show photon anticorrelation or photon antibunching. Whereas a group of photons or multiphoton emission can be detected simultaneously at two different locations and will show a correlation or photon bunching effect. These effects are measured using Hanbury-Brown-Twiss interferometer (HBT) setup[22]. Generally, the photon antibunching is reported in terms of the normalized second order correlation at a lag time 'Zero' ($g^2(0)$) value, which indicates the fraction of multi photon emission from an SPS. The minimum value of $g^2(0)$ represents the increase in SPS efficiency, and a high photon purity can be achieved with the lower $g^2(0)$ value. In addition to the above, the increased photon count rate by decreasing the fluorescence blinking events along with the increased "ON time" can provide another dimension for the efficient SPS at room temperature.

Recently, CdSe/CdS nanorods,[23] InP/ZnSe QDs[24] and PbX QDs[20] have shown promising results as efficient SPS at room temperature. Inspired by the above results, here, we present a new series of water soluble red emissive easily synthesized $CdTe_xSe_{1-x}$ QDs with varying composition of Se (Selenium), for the efficient single photon emission at room temperature. Our finding revealed an increase in single photon purity and deterministic emission with Se alloying and it was found to be maximum in $CdTe_{0.25}Se_{0.75}$. It showed increased photon counts rate, substantially reduced blinking events and increased single particle "ON-time". Previous reports indicated the importance of Auger recombination[20,24,25] and quantum confinement in determining the fate of QD's single photon purity. However, the increased single photon emission, in the present case, can be attributed to very fast electron trapping to very dense trap states with suppression of multiexcitons, leading to single photon emission and enhanced fluorescence ON times from the QDs.

**Results and Discussion**

Varied composition of water soluble $CdTe_{1.00}Se_{0.00}$ (**QD1**), $CdTe_{0.75}Se_{0.25}$ QDs (**QD2**), $CdTe_{0.50}Se_{0.50}$ QDs (**QD3**) and $CdTe_{0.25}Se_{0.75}$ QDs (**QD4**) were synthesized by using cadmium chloride, sodium tellurite and sodium selenite as precursor molecules. Sodium borohydride and Hydrazine hydrate were used as reducing agents, and mercaptosuccinic acid as a capping agent. The detailed synthesis protocol is available in the supplementary information (SI). A thorough characterization of all the QDs were carried out to understand the materials property. Powder XRD patterns (**Fig. 1a**) show mainly three 2θ peaks at 23.88°, 39.61°, and 46.62°. These peaks correspond to (111), (220) and (311) planes of CdTe crystal structure. For QD1, the value aligns well with standard values for reported CdTe QDs[26]. A gradual shifting of these peaks can be clearly observed from 23.88° to 25.30° for (111), 39.72° to 41.87° for (220) and 46.43° to 49.85° for (311) for QD2 to QD4, thus suggesting the formation of varied Se composition in CdTe QDs. X-ray photoelectron spectroscopy (XPS) (**Fig. S1**) showed the presence of cadmium, tellurium, and sulphur in all QDs. Selenium is found to be absent in QD1 (**Fig. S1a inset**) and progressively increases from QD2 to QD4, confirming its presence in these QDs. High resolution (HR) XPS (**Fig. S2 and Table S1**) show Cd $d_{5/2}$ & $d_{3/2}$ binding energy (BE) peaks at ~405 eV & ~412 eV, respectively, Te $d_{5/2}$ & $d_{3/2}$ BE peaks at ~572 eV & ~582 eV. Deconvolution for both peaks yielded two curves (Red and Blue in **Fig. S2**) for QD's core and surface Cd and Te atoms, since they have slightly different BE. Also, 572 eV peak for Te signifies the formation of $Te^{2-}$, from CdTe formation[27]. Se 3d peak at ~54.6 signifies the formation of $Se^{2-}$ from CdSe[28]. S 2P peaks at around 162 signifies the formation of sulphide ($S^{2-}$) and thiolate (Cd-S-R) at the surface of QDs[27]. Transmission electron microscopy (TEM) micrographs (**Fig. S3**) show the formation of QDs with crystallization of (111) plane with d-spacing of 3.22 Å for QD1 (**Fig. S4a**), which increased slightly to 3.42 Å for QD2 (**Fig. S4b**), 3.46 Å for QD3 (**Fig. S4c)** and 3.55 Å for QD4 (**Fig. S4d**). The d-spacing data was gathered from ~20 QDs and plotted in **Fig S5.** This change in d-spacing signifies the change in composition of individual QDs lattice structure due to selenium alloying[29,30]. From the TEM data, it is evident that all the QDs size lie in between 3.8-4.6 nm. Next, we carried out the optical characterization of these materials. Transient absorption spectroscopy (TAS) was performed by pumping the QDs with ~380 nm pulsed light source and probed with white light. TA Spectral analysis showed a single bleach signal at ~600 nm for QD1, which is redshifted from QD2 to QD4 (**Fig 1b and Fig. S6**). Interestingly, a new peak at ~520 nm gradually evolved from QD1 to QD4. All these characterization techniques clearly signify the formation of varied compositions of alloyed QDs. UV-Visible absorption spectroscopy (**Fig. 1c**) revealed band edge absorption at 587 nm, 598 nm, 619 nm, and 648 nm for QD1, QD2, QD3, and QD4, respectively. All QDs showed strong absorption below 400 nm, which exists due to a higher density of states in valence and conduction bands at the continuum level. Photoluminescence (PL) spectrum (**Fig. 1d**) showed emission maxima shifting from 604 nm in QD1 to 617 nm in QD2, 641 nm in QD3 and 684 nm in QD4. The red shifting in absorption and PL spectrum signifies that the band edge transition energy decreases due to alloy formation.

For utilizing QDs as an efficient SPS, the QD must exhibit deterministic emission or the on-demand emission of single photons. Hence, it is essential to study the single particle level fluorescence of QDs individually. For this, a very dilute sample of QDs was spin coated over a cleaned glass coverslip. The individual QDs were observed under continuous wave (CW) mode of excitation with single photon sensitivity of electron multiplying charge coupled device (EMCCD). Detailed information can be found in the SI. Nearly 400 QDs for each sample were analyzed, and the data is represented in **Fig. 2** and **Fig S7**. **Fig. 2a-d** showed the representative real time traces of single QD1 (**Fig. 2a**), QD2 (**Fig. 2b**), QD3 (**Fig. 2c**), and QD4 (**Fig. 2d**). Se alloying increased the mean total fluorescence ON time from ~4 s in QD1, to ~30-50 s in QD3 and QD4 (**Fig. 2e**). We also observed a few QD3 and QD4 with total ON time greater than 240 s which signifies very high deterministic emission (more than 95% ON time) of QD3 and QD4. However, total OFF time displayed no correlation with the Se alloying (**Fig. 2f**). We also found that the mean photon count rate or fluorescence intensity increased from ~ counts ~5000 in QD1, ~6500 in QD2, ~14000 in QD3, and ~18000 in QD4. The increment in intensity and the ON times clearly advocate for the bright and deterministic emission from QD3 and QD4 in comparison to QD1 and QD2.

Getting inspired by the above results, we next carried out the fluorescence antibunching experiment using a classical HBT setup to check the single photon purity experiments of these QDs (**Fig. 3a**) (detailed experimental conditions in SI). In brief, the diluted QDs sample was illuminated with a CW laser of 488 nm with a 60x water immersion objective. The emission of QDs was collected and sent to the HBT setup, which consists of two single photon avalanche photodiode (SPAD) detectors kept perpendicular to each other, and the QDs emitted light is passed through a 50:50 beam splitter. This beam splitter splits the light to two detectors and a second order correlation ($g^2(\tau)$) is generated. A sharp dip can be seen at zero-time delay, as can be seen in **Fig. 3b**. This sharp dip signifies photon antibunching and ensures single photon emission from all QDs. It was determined that the average $g^2(0)$ value was 0.22 in QD1 and then gradually decreased to 0.19 in QD2, 0.1 in QD3, and finally to 0.02 in QD4 (**Fig. 3c**). It is noteworthy to be pointed out that the selenium alloying enhanced the single photon purity of QDs significantly and QD4 displayed ultra-high single photon purity in comparison to other QDs.

A clear understanding of the photon emission and recombination pathways in QDs is essential for optimizing their performance as an SPS. By understanding the mechanism of QD emission, we can better control and optimize the factors that influence the photon emission efficiency of QDs in quantum technologies to develop efficient SPS. TAS results were utilized to understand the detailed mechanism of the SPS of these QDs. The decay kinetics analysis was performed with QDs pumped at 380 nm and probed at the ground state bleach signal (600 nm for QD1, 620 nm at QD2, 650 nm at QD3 and 690 nm at QD4). The relaxation or cooling time for charge carriers is shown in **Fig. 4**. Cooling time for QD1 was found to be 295 fs, whereas for QD2 it was 342 fs, QD3 was 461 fs, and QD4 was 590 fs. Charge carriers in QD4 took more than twice the time to relax from pump excitation level to the band edge level in comparison to QD. An increase in trap state density may lead to increased cooling time for the excited state charge carrier as these trap states hinder the relaxation process[31].

Also, to neglect the changes caused by trap states in between pump and probe, we performed the excitation power dependent TAS decay kinetics experiments with pump and probe wavelengths kept close to each other (**Fig. 5**). In all QDs, a very long (slow) component exists, which can be attributed to the radiative recombination of the charge carriers. Also, we found that there exists a very fast component (15-20 ps) in selenium doped QDs (**Fig. 5b-d**), which was absent in QD1 (**Fig. 5a**). However, this kind of fast component is not observed in the PL lifetime of selenium alloyed QDs, as observed in **Fig. S8.** Both observations suggest a very fast non-radiative recombination process occurring in the Selenium alloyed QDs (QD2-QD4), in comparison to the non-alloyed QD1. These very fast time scale components are generally attributed to the electron trapping[32]. It should be noted that even on higher intensity (power) excitations, the time scales for this fast non radiative component did not change in QD2-QD4, suggesting excitation power independence of this fast component. Whereas TA decay kinetics of QD1 exhibited strong power dependence. Such a power dependence may exist due to the presence of multi-exciton recombination or auger recombination, since these processes are dependent on the charge carrier concentration. Also, electron trapping process is absent in QD1 as the very fast component (<20 ps) was absent and instead, a short component (70-100 ps) is present in it, which can be attributed to the multi exciton recombination and auger recombination processes.

**Scheme 1** illustrates the mechanism of PL emission, and single photon versus multiphoton emission. Upon excitation with a light source, electrons are excited, forming single or multiple excitons depending on the absorption process. These electrons then relax to the band edge state. CdTe QDs cool faster (295 fs), whereas $CdTe_{0.25}Se_{0.75}$ takes longer time to cool, indicating higher trap state density in $CdTe_{0.25}Se_{0.75}$ QDs than CdTe QDs. This higher density of trap states facilitates the very fast electron trapping, typically within 20 ps. The electron trapping occurs faster than the multi exciton recombination processes. In CdTe QDs, the multi exciton recombination results in both multiphoton emission and single photon emission. On the other hand, in $CdTe_{0.25}Se_{0.75}$ QDs, the dominating contribution of electro trapping, suppresses the multi exciton, leading to decreased contribution of multiphoton emission. In QD1, both multi photon and single photon emission occur simultaneously, with a higher contribution from multiphoton emission, leading to a significantly higher $g^2(0)$ value. On the other hand, in QD4, due to the presence of higher density of trap states, charge trapping is highly feasible and occurs instantly in the time scale of few ps. Charge trapping rate is significantly higher in multi excitons in comparison to the single exciton and hence the single exciton recombination is effectively affected by increase in trap state density. This enhanced charge trapping reduces the efficiency of multi photon emission, ultimately increasing the photon antibunching with $g^2(0)$ value reducing to ~0.02.

**Conclusion**

We have presented $CdTe_{0.25}Se_{0.75}$ as one of the most promising materials for room temperature single photon emission source for quantum technology. Our study revealed an increase in fluorescence ON times and deterministic emission on increasing the Selenium composition in the core of quantum dots. This subsequent enhanced single photon emission with a very high fluorescence anticorrelation, $g^2(0)$ (~0.02) is attributed to very fast electron trapping to very dense trap states with suppression of multiexcitons. Our findings also revealed that quantum confinement and Auger recombination are not the sole factor in determining the fate of single photon emission or $g^2(0)$ values in QDs. These findings may support future developments of SPS in the field of quantum computing by designing quantum dots with engineered trap states and auger rates.

## Conflict of Interest

The authors declare no conflict of interest.

## Author Contributions

KK conceptualized and designed all the experiments. KK, JM, and RKB performed all the bulk-level measurements. KK performed single particle spectroscopy measurements. SS performed the XPS analysis. KK analyzed the data with the help of CKN. CKN guided the complete project and completed the writing of the manuscript with the help of KK.

## Acknowledgements

CKN is thankful to Science and Engineering Research Board (SERB) core research grant (CRG) India for the project number CRG/2020/000268. CKN is thankful to the facilities of the AMRC centre of IIT Mandi, India. KK and CKN acknowledge the Sophisticated Analytical and Technical Help Institutes (SATHI), IIT Delhi and Indian Science Technology and Engineering facilities Map (I-STEM) for fluorescence antibunching measurements. KK and CKN also thank Prof. Suman K. Pal, School of Physical Sciences, IIT Mandi for Transient Absorption Spectroscopy measurements. Graphical abstract and schematics were created using Biorender.com.

## Corresponding author information

Name: Chayan Kanti Nandi

Designation: Professor

Email: chayan@iitmandi.ac.in

ORCID: https://orcid.org/0000-0002-4584-0738

Address: A3 Academic Block, School of Chemical Sciences, South Campus, IIT Mandi, H.P. 175075

**Manuscript Figures content**

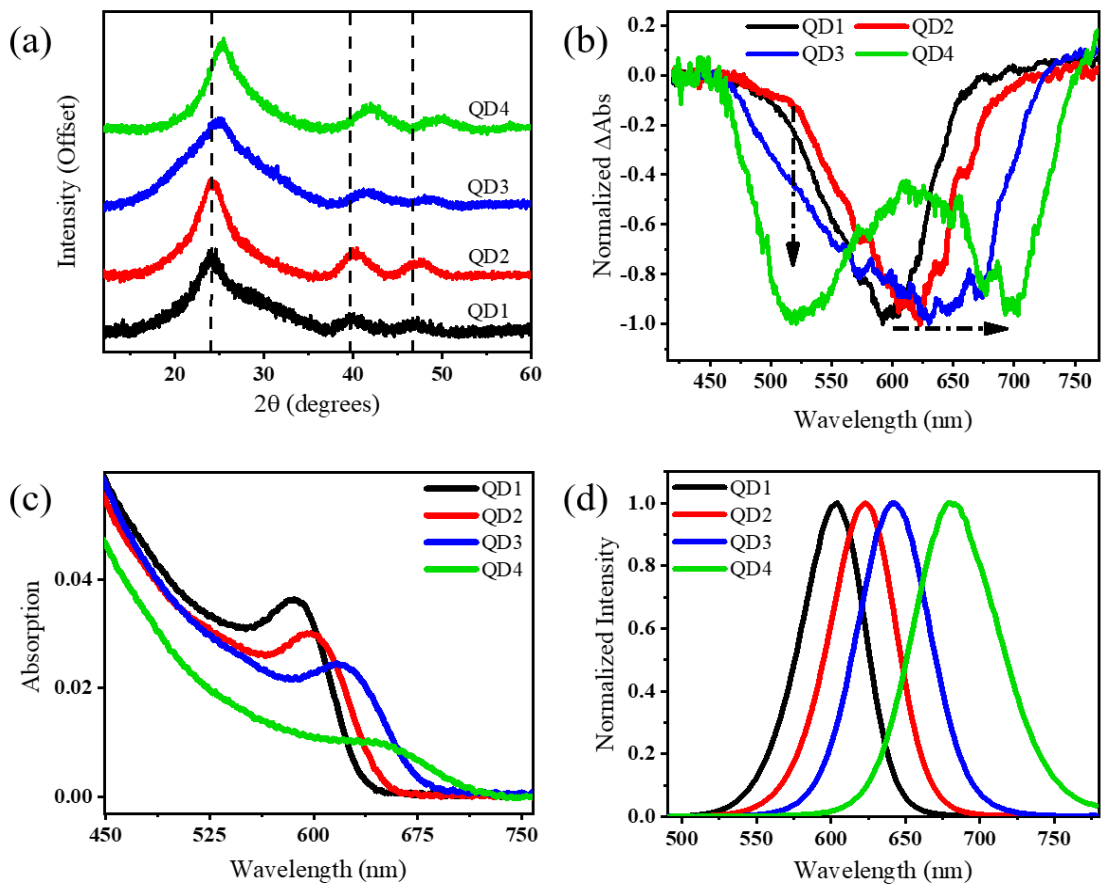

**Fig. 1: Structural and photophysical characterization of synthesized QDs**: (a) XRD patterns show shifting in 2θ (degrees) from QD1 to QD4. Vertical dashed lines correspond to the peak for CdTe QDs at 23.88°, 39.61°, and 46.62°. (b) Transient absorption spectrum of QD1-QD4 pumped with 380 nm pulse and probed with white light. Selenium alloying decreases the band gap and a shift in 600 nm peak is observed. Also, a new peak evolved at ~520 nm. (c) UV-Visible absorption spectrum with band edge absorption peak for CdTe QDs (**QD1**) at ~587 nm, $CdSe_{0.25}Te_{0.75}$ QDs (**QD2**) at ~598 nm, $CdSe_{0.50}Te_{0.50}$ QDs (**QD3**) at 619 nm and $CdSe_{0.75}Te_{0.25}$ QDs (**QD4**) at 648 nm. (d) PL Spectra show red shifting from 604 nm in QD1, ~617 nm for QD2, 641 nm in QD3, and 684 nm in QD4.

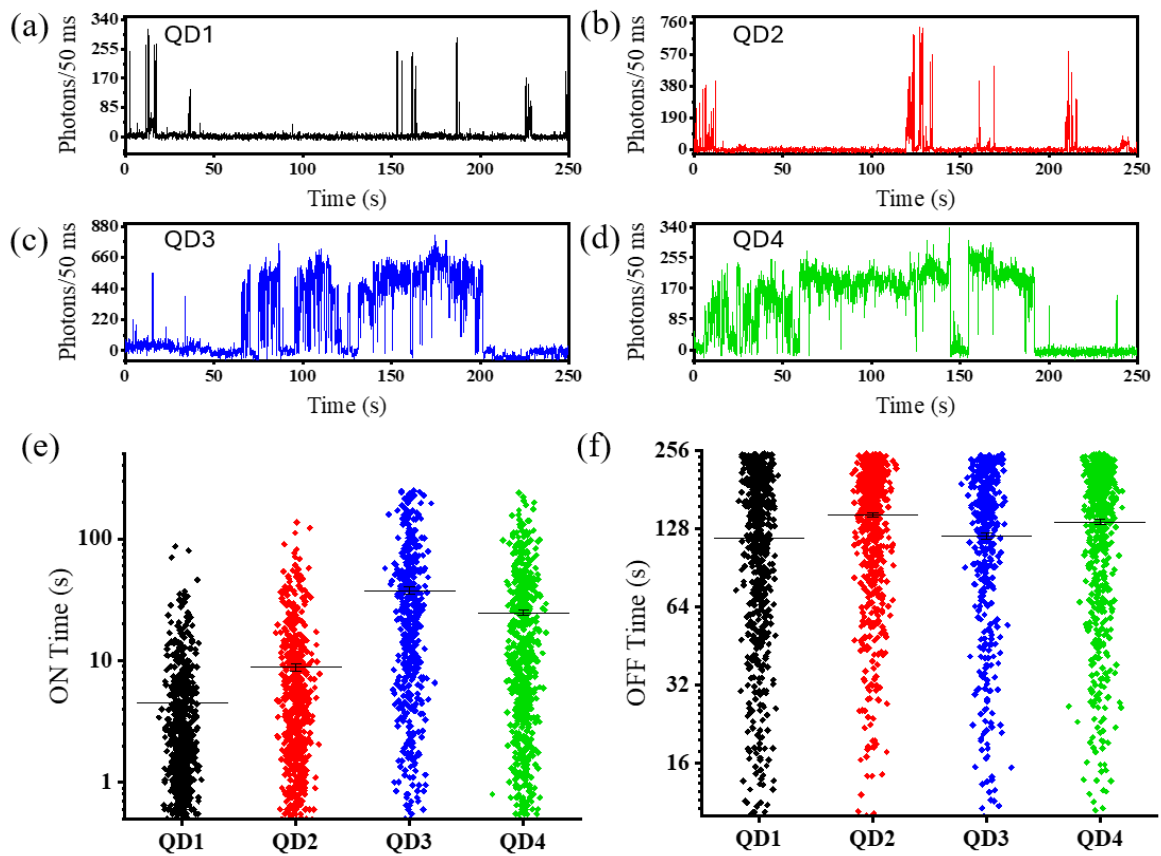

**Fig. 2: Single particle fluorescence spectroscopy**: Representative single particle fluorescence intensity-time traces of (a) QD1, (b) QD2, (c) QD3, and (d) QD4. The data show very sharp photon bursts in QD1 and QD2. Whereas QD3 and QD4 have significantly higher ON times. ~400 of such individual QD-time traces were analyzed and (e) Total ON time from all QDs plotted together. (f) Total OFF time from all QDs plotted together. The horizontal line in e & f represents the mean value of all represented data. Few of the representative time traces are given in **Fig. S7**

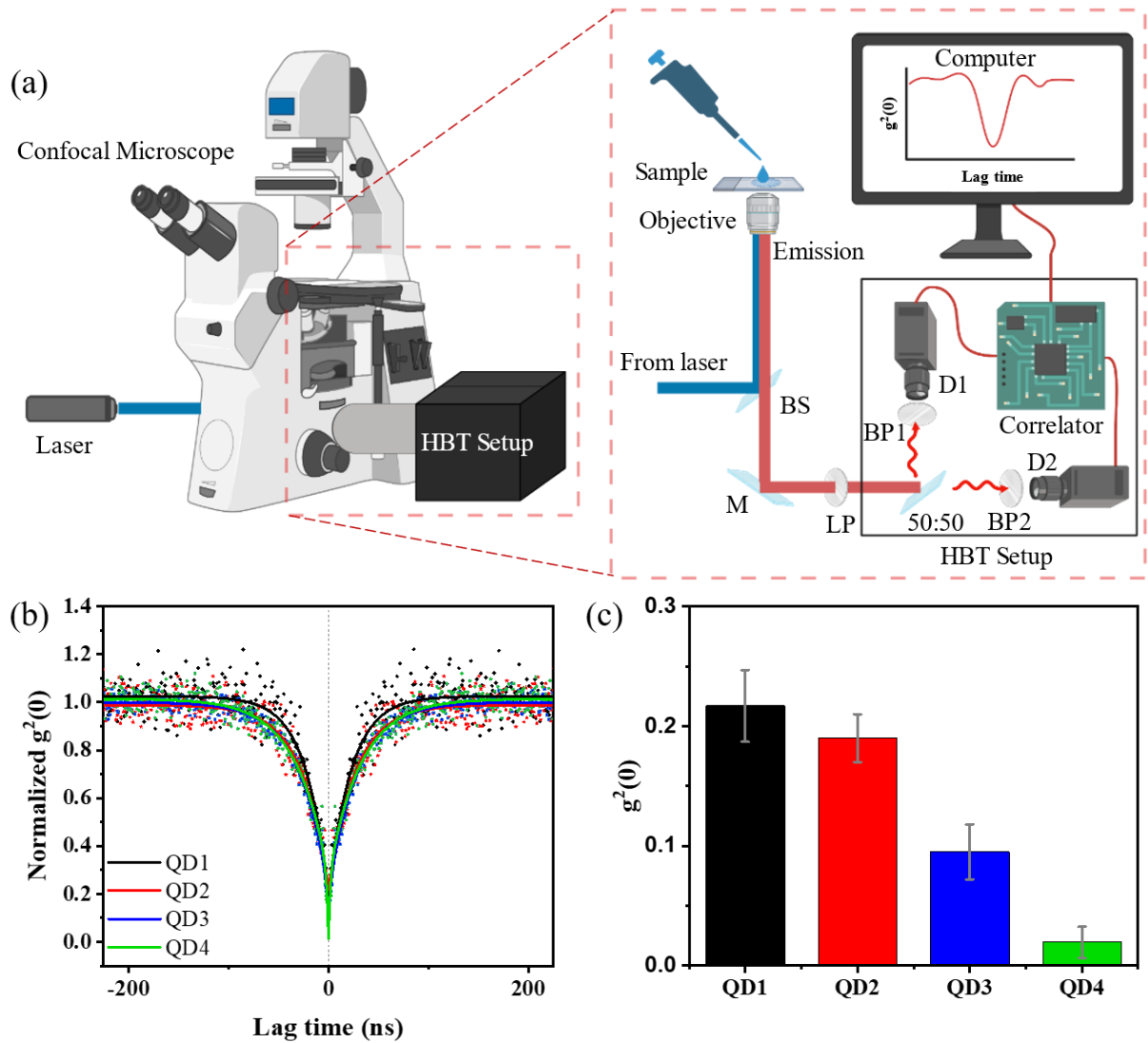

**Fig. 3:** (a) Experimental setup for determining second order photon correlation or fluorescence antibunching using Hanbury Brown Twiss (HBT) setup: CW laser excites sample placed over a confocal microscope. Emitted light from sample reaches the HBT setup via beam splitter (BS), Mirror (M), a long wavelength pass filter (LP). In HBT Setup, the emission light from the sample is split into two directions (one transmitted and one reflected) via a 50:50 beam splitter. Two same band pass filters (BP1 & BP2) were placed in front of each detector to avoid the detector afterglow. A correlator performs the real time correlation of incident photons at detector D1 and D2 (SPADs) and in case of single photon emission, both D1 and D2 shows high correlation and for the case of single photon emission, single photon can either be detected at detector D1 or D2 and correlator determines no correlation as can be seen in the dip at time 'zero'. (b) Normalized fluorescence antibunching data showing a sharp dip in all the QDs near lag time 'Zero'. The data was fitted with two component exponential equation. (c) comparison of $g^2(0)$ value obtained from the extrapolation of the exponential fitting the data in (b) showing increasing single photon emission from QD1 to QD4.

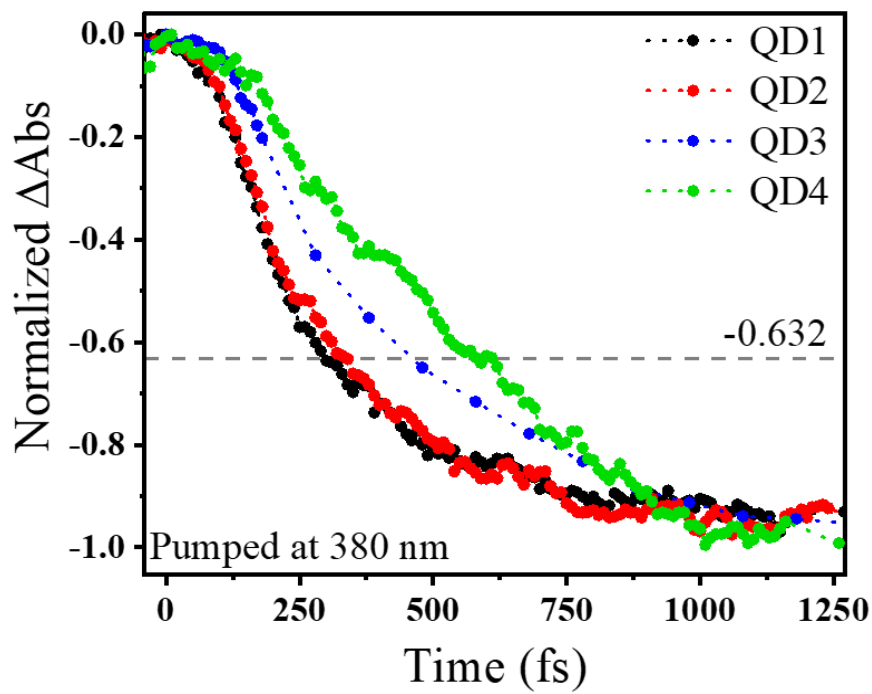

**Fig. 4:** Transient absorption decay kinetics recorded with QDs pumped with 380 nm laser and probed near the band edge of QDs i.e. 600 nm for QD1, 620 nm for QD2, 650 nm for QD3 and 690 nm for QD4. The vertical dashed line represents the 1/e of the initial Normalized ΔAbs value. Cooling time is ~295 fs in QD1, ~342 fs in QD2, ~461 fs in QD3 and ~590fs in QD4. A clear delay in cooling time is observed in Selenium alloyed QDs.

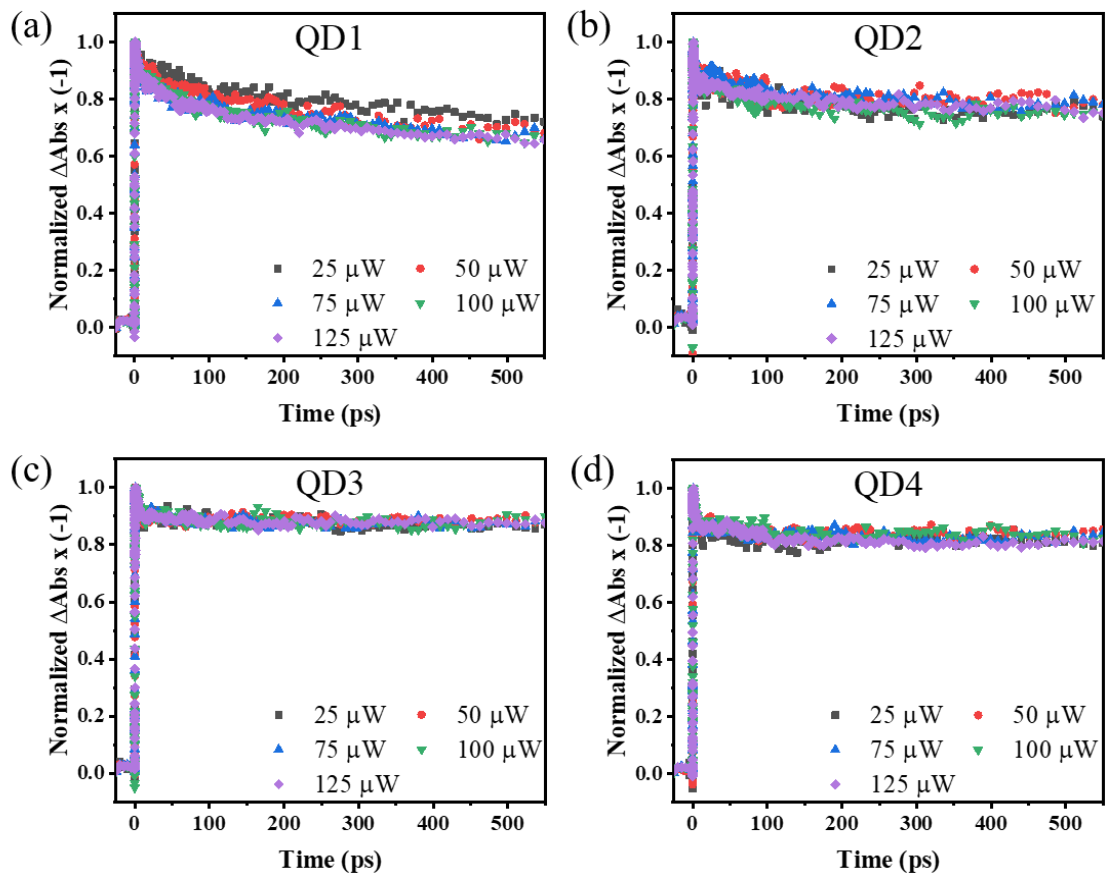

**Fig. 5:** Excitation power dependent transient absorption decay kinetics analysis for QDs pumped and probed near the band edge states: (a) QD1, (b) QD2, (c) QD3, and (d) QD4. A clear power dependence can only be observed in QD1. QD2-QD4 exhibits no power dependence.

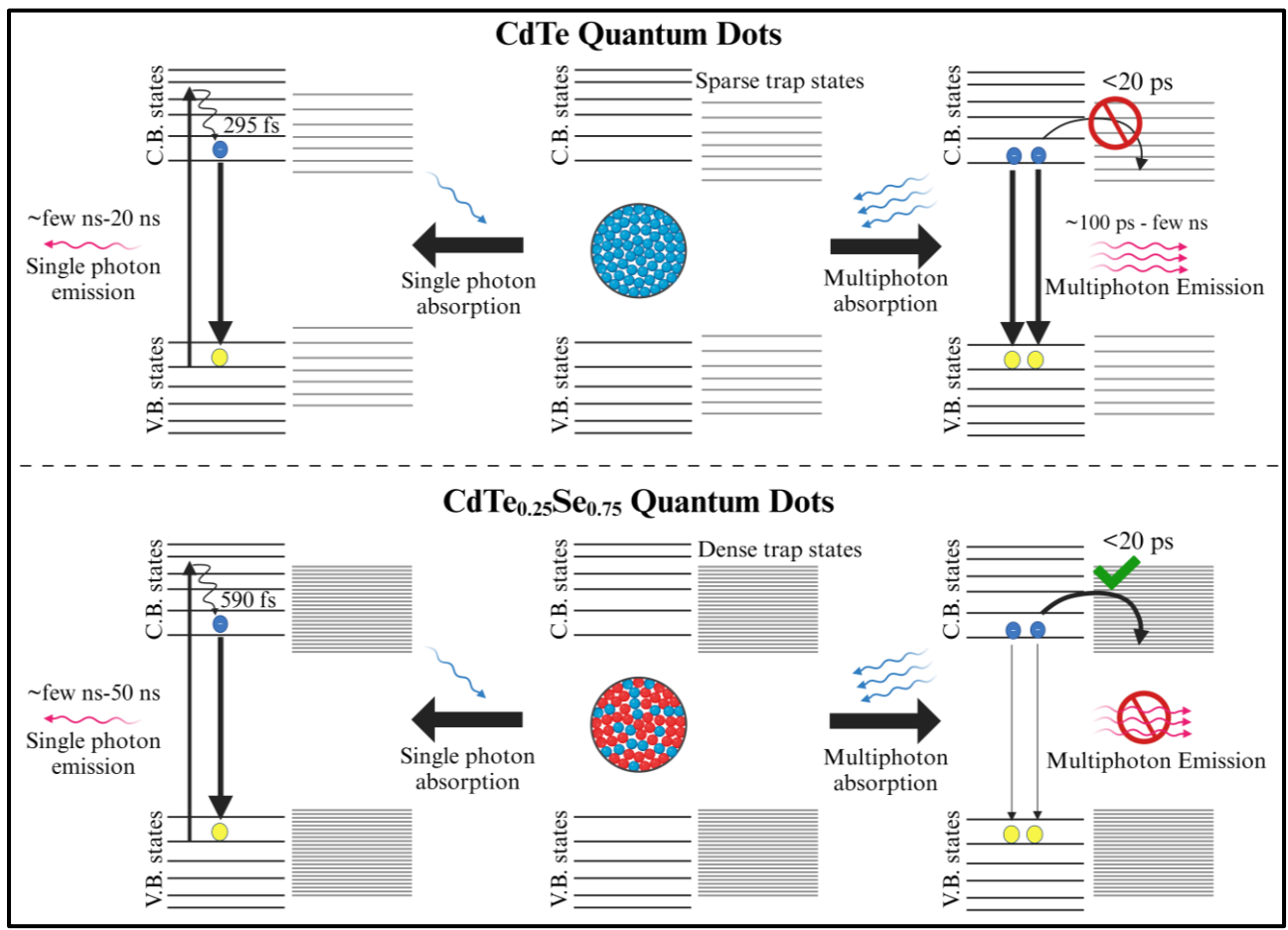

**Scheme 1: Schematic illustrating the mechanism for enhanced single photon emission from** CdTe$_{1.00}$Se$_{0.00}$ (**QD1**), CdTe$_{0.25}$Se$_{0.75}$ QDs (**QD4**). Due to the increased density of states, the single photon emission probability increases in CdTe$_{0.25}$Se$_{0.75}$ QDs (**QD4**)

**Graphical Abstract**

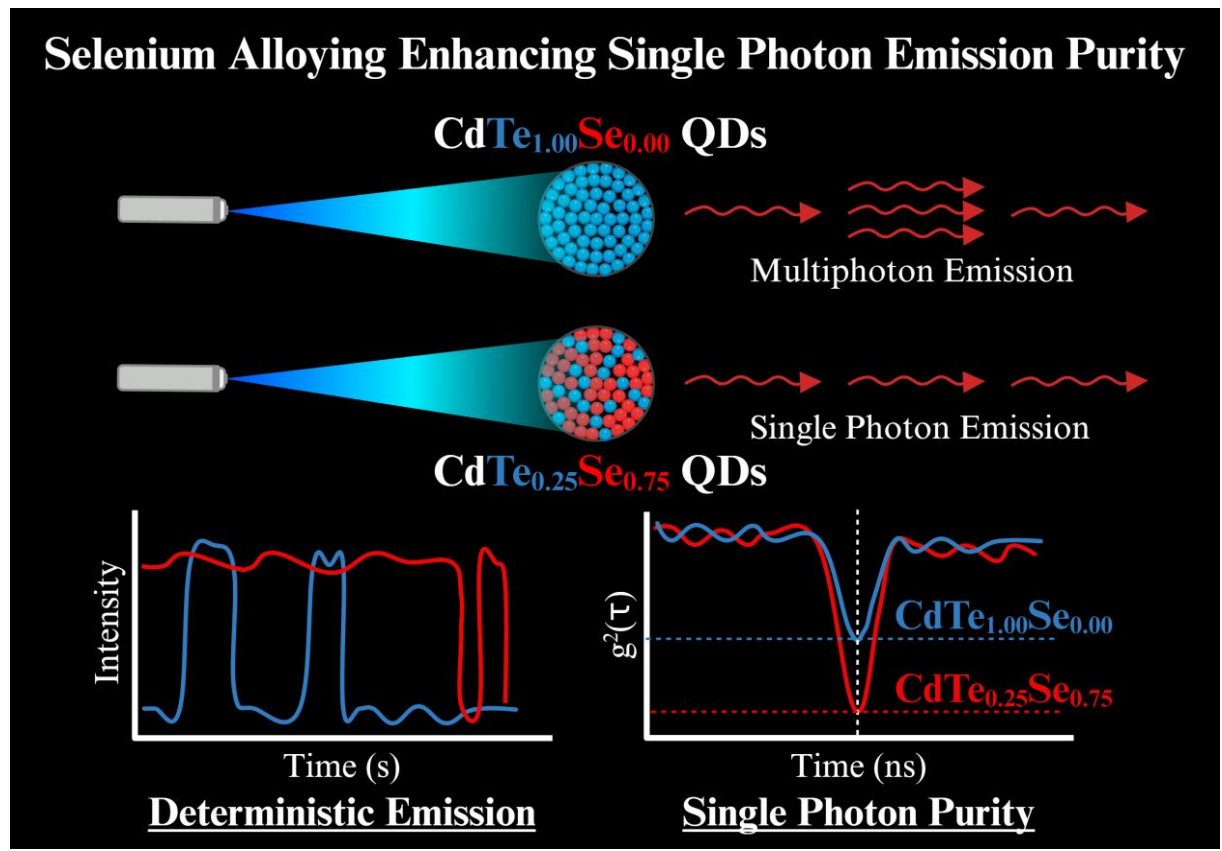

CdTe$_{0.25}$Se$_{0.75}$ quantum dots shows substantial enhancement in single photon emission purity

# Electronic Supplementary information

## Experimental Section

### Materials

Cadmium chloride hemipentahydrate $CdCl_2.2.5H_2O$, Mercaptosuccinic acid (MSA), Sodium selenite was purchased from Sigma. Sodium tellurite was purchased from Alfa-Aesar. Sodium Chloride was purchased from Merck. Sodium borohydride was purchased from Avra. Hydrazine hydrate 80% was purchased from Loba Chemie private limited. Sulphuric acid and 30% hydrogen peroxide were purchased from fisher scientific. All the commercial samples were used without further purification.

### Synthesis of water-soluble alloyed quantum dots (QDs)

The QDs were synthesized as follows: First, 40 mL of deionized water was deaerated for approximately 30-45 minutes using nitrogen gas purging. Then, 0.3425 g (1.5 mmol) of $CdCl_2 \cdot 2.5\ H_2O$ and 0.2477 g (1.65 mmol) of MSA were dissolved in the deaerated deionized water. The pH of this solution was adjusted to between 8 and 10, and it was labeled as Solution A. Solution A was then divided into five parts, each containing 8 mL, and placed into 15 mL test tubes labeled A1 to A5. Separately, 10 mL solutions containing 0.25 mmol of sodium tellurite (Solution M) and sodium selenite (Solution N) were prepared. A batch of four solutions with varying ratios of M:N (2 mL:0 mL for B1, 1.5 mL:0.5 mL for B2, 1 mL:1 mL for B3, and 0.5 mL:1.5 mL for B4) were created in 2 mL microcentrifuge tubes.

Next, a set of solutions were made by combining each Ai solution (A1 to A4) with the corresponding Bi solution (B1 to B4) and adding 13.2 mg of sodium borohydride to each solution, resulting in solutions labeled C1 to C4. Rice beads were added to each test tube to ensure better mixing. On continuous stirring, solution C1 turned deep yellow (after approximately 20 minutes), then 400 µL of $N_2H_4$ was added to all solutions. The solutions were then kept in a 95 °C water bath for 30 minutes. Resulting in solution QD1, QD2, QD3, QD4. These solutions were further dialyzed for purification and used for further studies.

### UV Visible absorption spectroscopy

Shimadzu UV 2540 spectrophotometer was used to record the absorption spectrum for all QDs. Baseline was taken before all acquisitions using water. 10 mm path length 1 mL quartz cuvettes were used at both sample and reference positions. UV probe software was used for acquisition.

**Photoluminescence (PL) spectroscopy**

Steady state PL spectrum was recorded using Agilent technologies Cary Eclipse fluorescence spectrophotometer. 10 mm path length, 1 mL quartz cuvette was used. Agilent Cary WinFLR fluorescence software was used for data acquisition and analysis.

**PL lifetime decay spectroscopy**

The PL lifetime decay was recorded using the Horiba Scientific Delta Flex TCSPC system with 454 nm pulsed LED source. The emission monochromator was set to the emission maxima position of the QD sample. Ludox was used for IRF determination. Horiba's Data Station was used for acquisition and DAS6 and Origin 2018 were used for analysis.

**Transmission electron microscopy (TEM)**

FEI Tecnai, USA, FP 5022/22-Tecnai G2 20 S-Twin, operating at 200 keV was used for TEM imaging of all QDs. Obtained images were further analyzed using Gatan Digital Micrograph software for d-spacing calculation.

**Transient absorption spectroscopy**

A Ti:sapphire laser amplifier (Spitfire Ace, Spectra Physics) was employed for ultrafast transient absorption (TA) measurements. The pump and probe pulses were obtained from the amplifier by splitting the output beam pulse width < 35 fs and wavelength~800 nm) into two components. The pump and probe pulses of different energies (used in single kinetics measurements) were obtained from the automated optical parametric amplifier (TOPAS prime from Spectra Physics). In broadband TA measurement, the probe beam was a white-light continuum (WLC) generated by focusing a small fraction of 800 nm light on a sapphire crystal. A delay stage was used in the probe path to control the time delay between the probe and pump beams. The absorbance of the probe beam was detected under the conditions with and without pump with the help of a mechanical chopper. TA spectra were recorded by dispersing the beam with a grating spectrograph (Acton Spectra Pro SP 2358) followed by a CCD array. Two photodiodes having variable gain were used to record TA kinetics. [Rephrase]

**X-ray photoelectron spectroscopy (XPS)**

QDs were drop casted on a Si wafer and dried for >12 hours. The Thermo Scientific NEXSA Surface Analysis instrument was used for the acquisition of XPS data. Avantage software was used to analyze the data.

**Single particle fluorescence spectroscopy**

Acquisition: Since the single particle experiments are very sensitive to the contamination, utmost care was taken to ensure that the glass coverslips were clean. Glass coverslips were treated with Piranha solution (3:1 solution of Sulphuric acid and Hydrogen peroxide) for half an hour, then the mixture was discarded, and glass coverslips were washed with ultrapure deionized water for 4 times and then were ultrasonicated. This process was repeated 4 times. Then the coverslips were kept in water till they were required for the experiment. Freshly cleaned glass coverslips were used for all the experiments. Two empty glass coverslips were analyzed for testing impurities to ensure cleanliness of glass coverslips.

QDs solutions were first diluted at nM to pM concentration, and the sample was spin-coated over a glass coverslip at ~5000 RPM (with 500 RPM/s acceleration). The coverslip was then mounted over a home-built inverted Nikon Ti epifluorescence inverted microscope objective 100x, 1.49 NA, TIRF objective. Laser beam of wavelength 488 nm was aligned for measurements. The laser beam reaches the glass coverslip using a 590 nm high pass dichroic mirror (F48-590 Beamsplitter (BS), AHF analysentechnik). After the sample is excited by these lasers, the emission is collected using the same objective. Then, the excitation and emission beams are separated by the same dichroic mirror, i.e., F48-590 BS. Emission is then further filtered using a appropriate band pass filter; 600/50 nm (F47-601, AHF analysentechnik) for QD1 and QD2 600/50 nm, 685/80 nm (F47-688, analysentechnik) for QD3 and 690/70 nm (F72-686, AHF analysentechnik) for QD4. Emission is finally collected at Andor EMCCD iXon Ultra 897.

Andor Solis 64-bit software was used to record the data. EMCCD was used in photon counting mode with an EM gain 300, exposure time 50 ms (~20 fps), and pixel readout rate 17 MHz. A movie with above mentioned settings, with 5000 frames (~250 s) was recorded and saved in FITS format. In the recorded video, one pixel corresponds to 160 nm x 160 nm area and an area of 20.48 µm x 20.48 µm (128 x 128 pixels) was recorded.

**Analysis of single particle spectroscopy data**

The area of the recorded video is 20.48 µm x 20.48 µm, and this huge area contains 30-70 numbers of QDs particles, and they show intensity fluctuations across the recorded video duration individually. We custom-built a script using ImageJ macro language to analyze the total photon counts and ON-OFF dwell times for 300-400 individual QDs. This script first identifies the bright QDs in maximum intensity projection (Z-project of ImageJ). Then a ROI box of 7 x 7 pixels is created around all of these identified localizations, and then the intensity vs. time graph is extracted for individual QDs. These obtained curves are single-particle raw time traces without any background subtraction.

Now a threshold is set up in all QDs to separate the ON and OFF states. Above the threshold, all emissions are considered ON and below a threshold value, all intensity fluctuations are OFF. All of these ON and OFF times are noted, and photon counts are also extracted from all such single-particle time traces. A fit line shows the ON states; for OFF states, the fit line reaches zero. It was ensured that no two ROI boxes overlap each other. In cases where ROI boxes overlap, all overlapping ROIs are discarded from the analysis. An ON time is considered as the time a QD particle spent in ON state without turning OFF. An OFF time is the time between two subsequent ON-states. If a QD turns dark and doesn't turn ON to the end of acquisition, that time is excluded from the OFF state since it is considered as photobleached or degraded.

All photon counts, and ON-OFF dwell times are then recorded in a separate Excel file and used for further analysis and plotting

**Fluorescence photon antibunching or fluorescence antibunching spectroscopy**

Fluorescence antibunching experiments were performed using Picoquant MicroTime 200. This instrument consists of PDL 828 Sepia II, MultiHarp 150, an Olympus inverted microscope with 60x water immersion objective and a laser combining unit with 405, 485, 532 nm pulsed diode laser. The MicroTime 200 system was used in T2 mode with sync killed. A drop of very diluted QDs sample (average number of particles < 0.6) was kept over glass coverslip. Then, the excitation laser (488 nm) in CW mode was focused on the sample, and emitted light was observed with two detector setups (both Excelitas single photon counting module SPCM AQRH single photon avalanche photodiode). A set of appropriate bandpass filters (two: one in front of each detector) were used in front of both detectors. Data acquisition was done for nearly one hour. The observed data was then analyzed using total correlation and antibunching analysis scripts in Symphotime 64 software. Data was normalized and then fitted using two-component exponential decay equation, i.e. $g^2(\tau) = y_0 + A_1 e^{\frac{-t}{\tau_1}} + A_2 e^{\frac{-t}{\tau_2}}$, using origin 2018 software. Here $\tau_1$ and $\tau_2$ represents the excitation time and relaxation times.

# Supplementary Figures contents

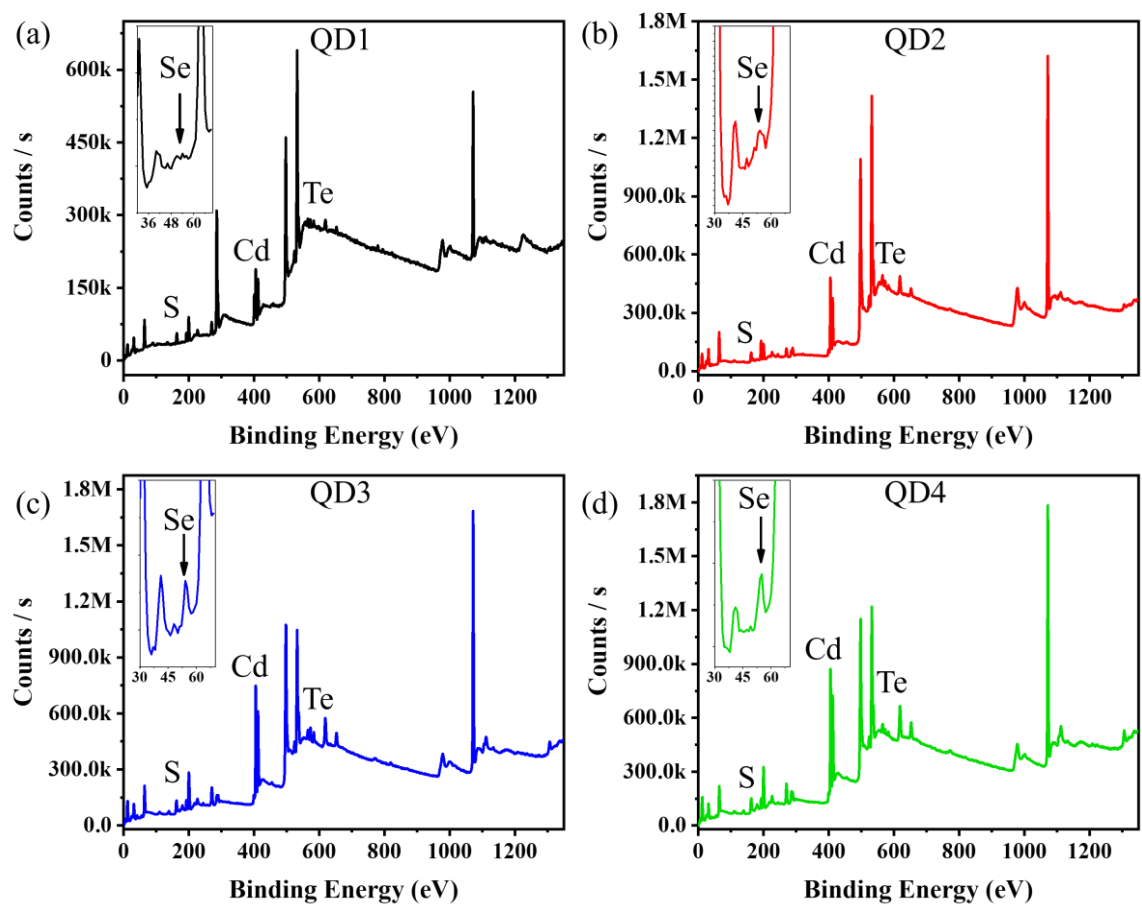

**Fig. S1:** XPS survey analysis of (a) QD1, (b) QD2, (c) QD3, and (d) QD4. There is a prominent with an increasing Selenium peak from QD2 to QD4, while it is absent in QD1. These results confirm the incorporation of Selenium in CdTe QDs

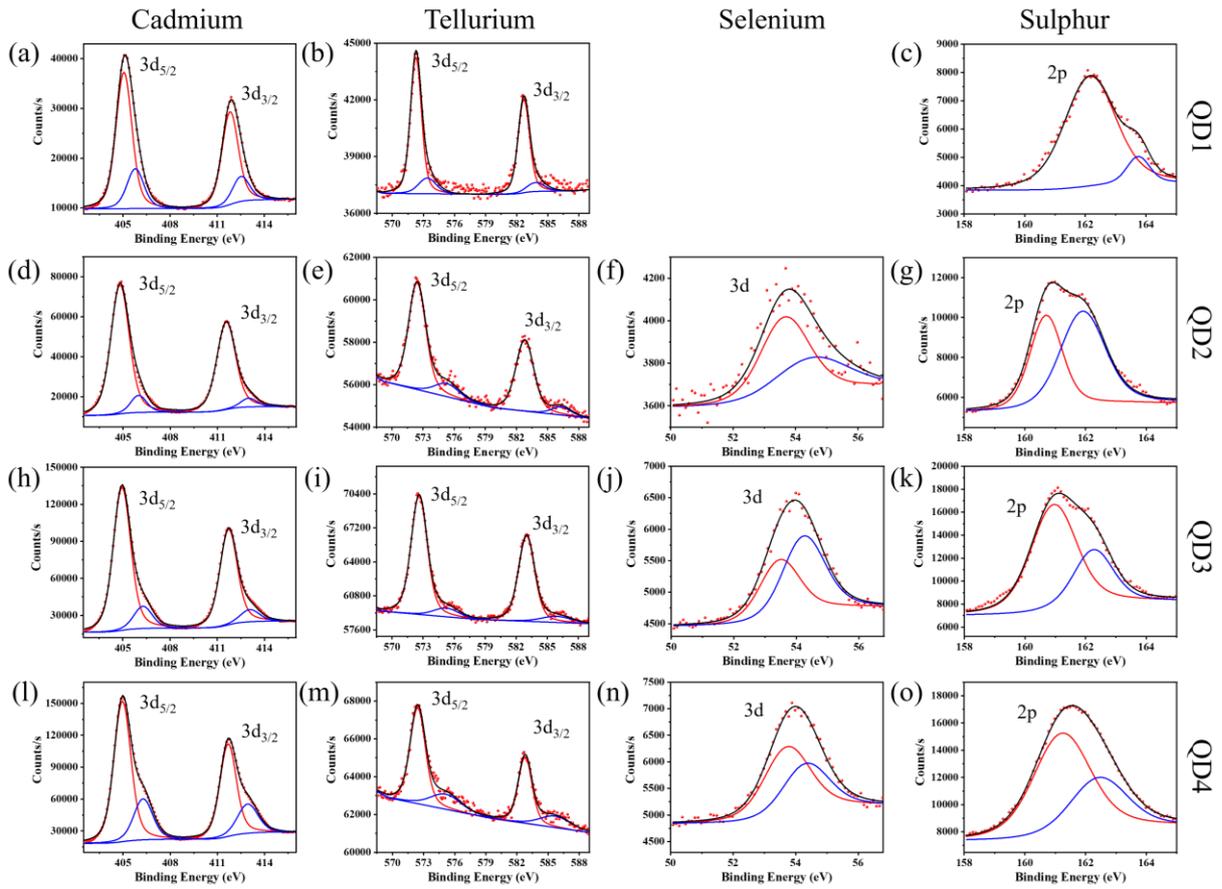

**Fig. S2:** Deconvolution of High-resolution XPS spectra of QD1 (a-c), QD2 (d-g), QD3 (h-k), QD4 (l-o) for the peaks of cadmium, tellurium, selenium, and sulphur. Cadmium and Tellurium are present in same oxidation states. Te ~572 eV peak shows Te²⁻, confirming presence of CdTe in all QD1-QD4. Se ~54.5 eV peak suggests the formation of Se²⁻ in QD2-QD4. Sulphur is present in all QDs as S²⁻ or Cd-S-R at the surface of QDs.

**Table S1:** Binding energy values (in eV) obtained from the deconvolution of high resolution XPS for major elements (Cd, Te, Se, S) present in the QD1-QD4.

| QDs | Deconvoluted Peak Color | Cd $3d_{5/2}$ | Cd $3d_{3/2}$ | Te $3d_{5/2}$ | Te $3d_{3/2}$ | Se 3d | S 2p |
|---|---|---|---|---|---|---|---|
| QD1 | Red | 405.08 | 411.78 | 572.28 | 582.68 | Absent | 162.18 |
| | Blue | 405.78 | 412.58 | 573.38 | 583.88 | | 163.78 |
| QD2 | Red | 404.78 | 411.58 | 572.38 | 582.78 | 53.68 | 160.68 |
| | Blue | 405.98 | 412.98 | 574.98 | 586.18 | 54.68 | 161.88 |
| QD3 | Red | 404.98 | 411.68 | 572.58 | 582.98 | 53.48 | 160.98 |
| | Blue | 406.28 | 413.08 | 575.08 | 585.78 | 54.28 | 162.28 |
| QD4 | Red | 404.98 | 411.68 | 572.38 | 582.78 | 53.78 | 161.28 |
| | Blue | 406.28 | 412.98 | 574.88 | 585.38 | 54.38 | 162.48 |

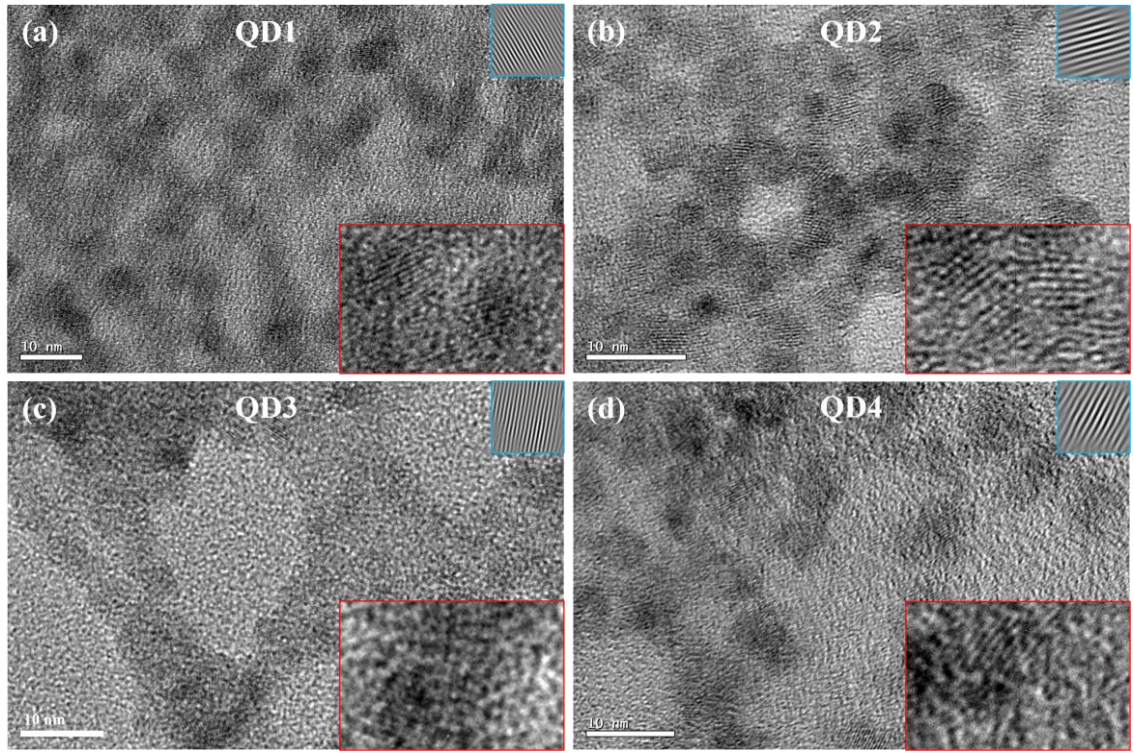

**Fig. S3:** TEM images for (a) QD1, (b) QD2, (c) QD3, and (d) QD4. QDs size lies in between 3.8 nm to 4.6 nm. The red border inset highlights the zoomed-in region of TEM image showing fringes in the core structure of QDs. Blue border in the inset shows fringe spacing obtained through FFT analysis, used further for d-spacing measurements.

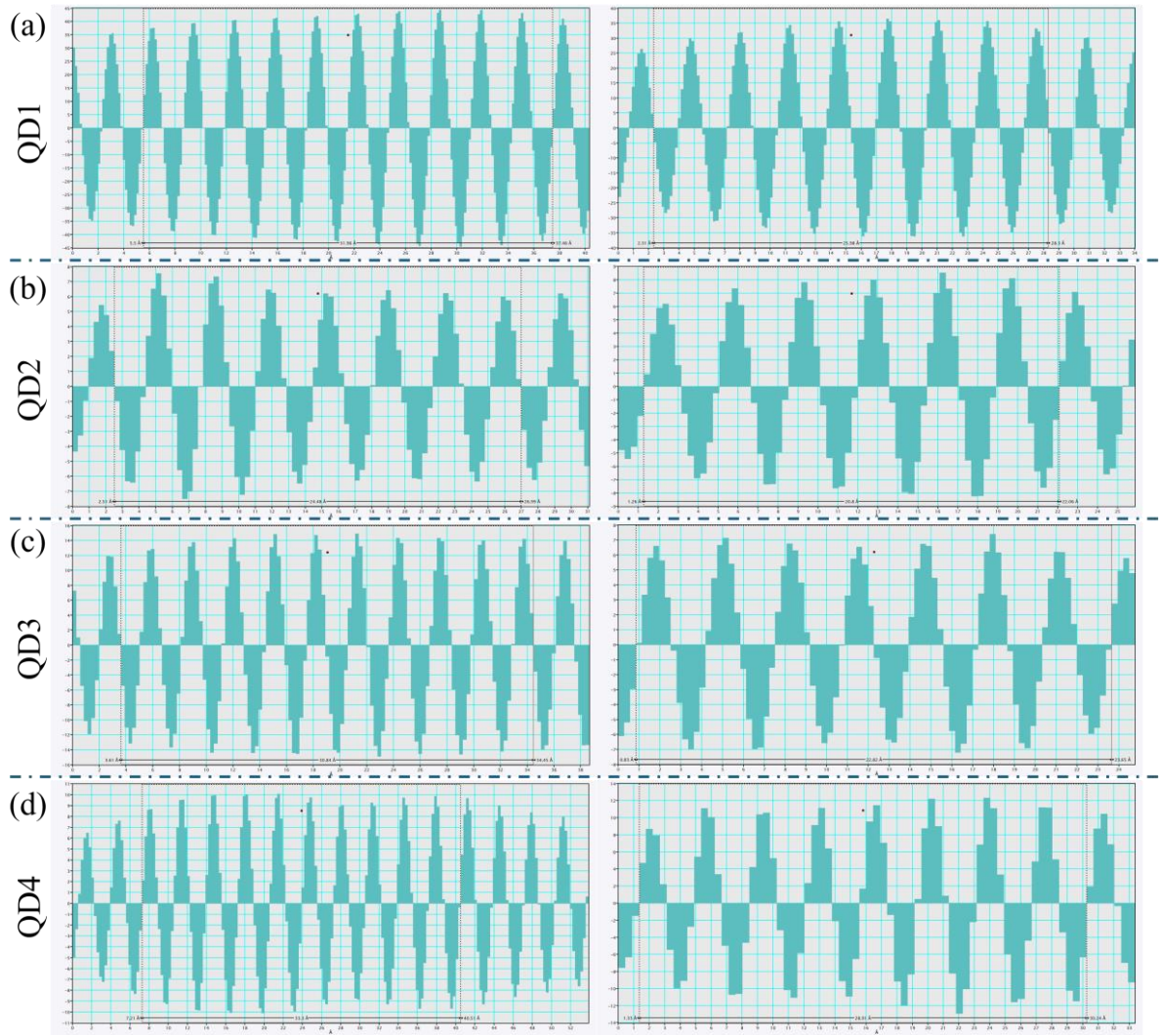

**Fig. S4:** Two of the representative line profiles are shown here for measuring the d-spacing of lattice fringes in HR-TEM of QDs, taken each from (a) QD1, (b) QD2, (c) QD3, and (d) QD4. x-axis unit is Angstrom.

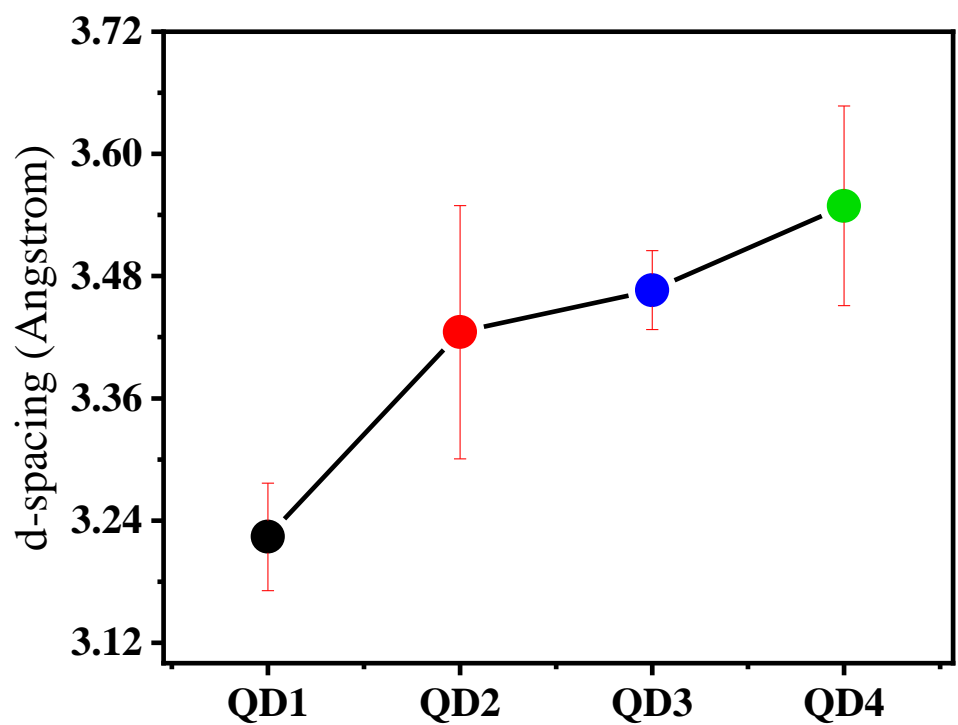

**Fig. S5:** d-spacing of QD1, QD2, QD3, and QD4 data collected from ~20 QDs. Circled data represents the mean d-spacing value of 3.22 Å for QD1, 3.42 Å for QD2, 3.46 Å for QD3, and 3.55 Å for QD4. Selenium alloying increases the lattice spacing in QDs core.

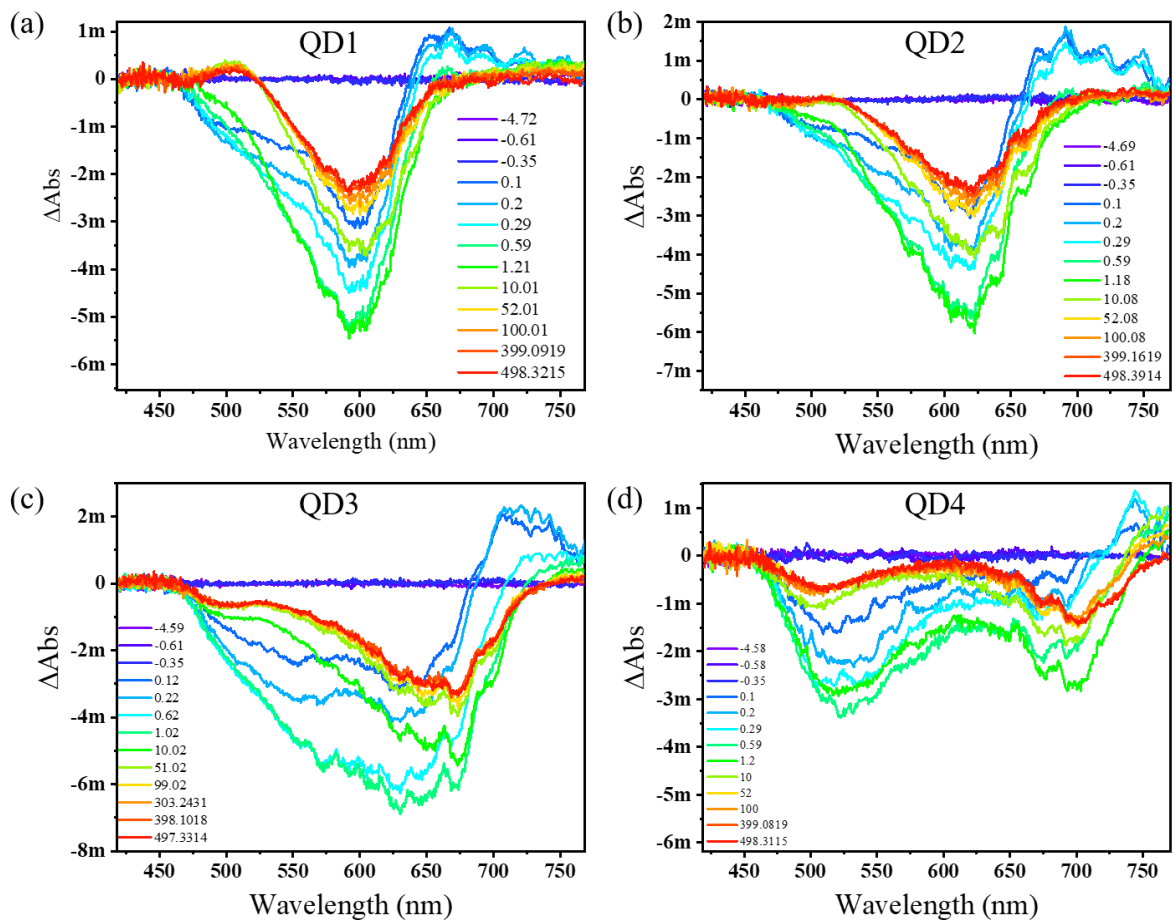

**Fig. S6:** Transient absorption spectrum for (a) QD1, (b) QD2, (c) QD3, and (d) QD4. Samples were pumped with 380 nm pulse and probed at different wavelengths using white light probe. A strong bleach signal appears at the band edge position which red shifts with selenium alloying. Also, a new peak evolved at ~520 nm in selenium alloyed QDs.

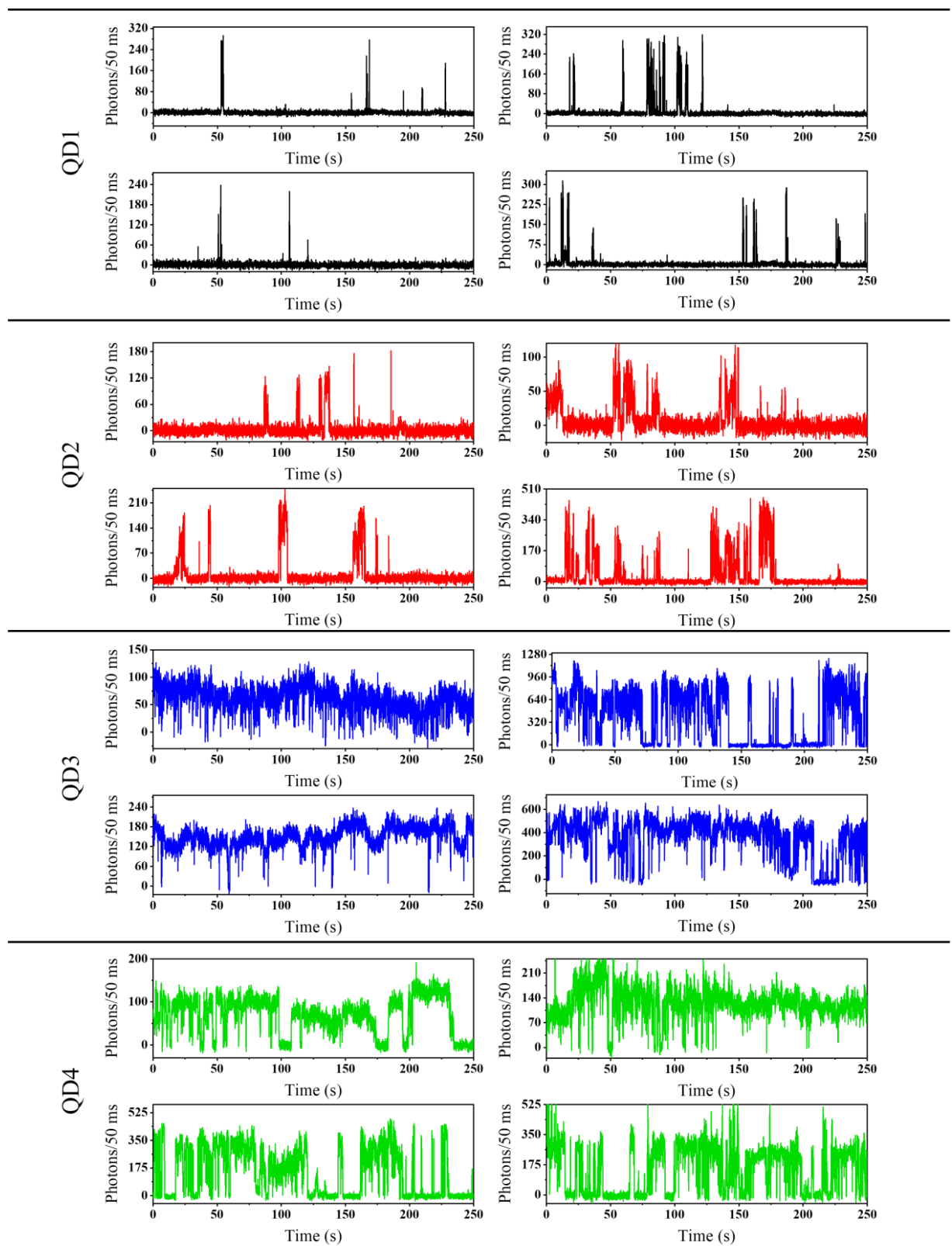

**Fig. S7:** Single particle fluorescence time traces for a few representative QDs. A clear increase in the fluorescence ON time is obtained from the QD1 to QD4 suggesting the increase in deterministic emission of selenium alloyed QDs. Few QD3 and QD4 also showed more than 95% of ON time.

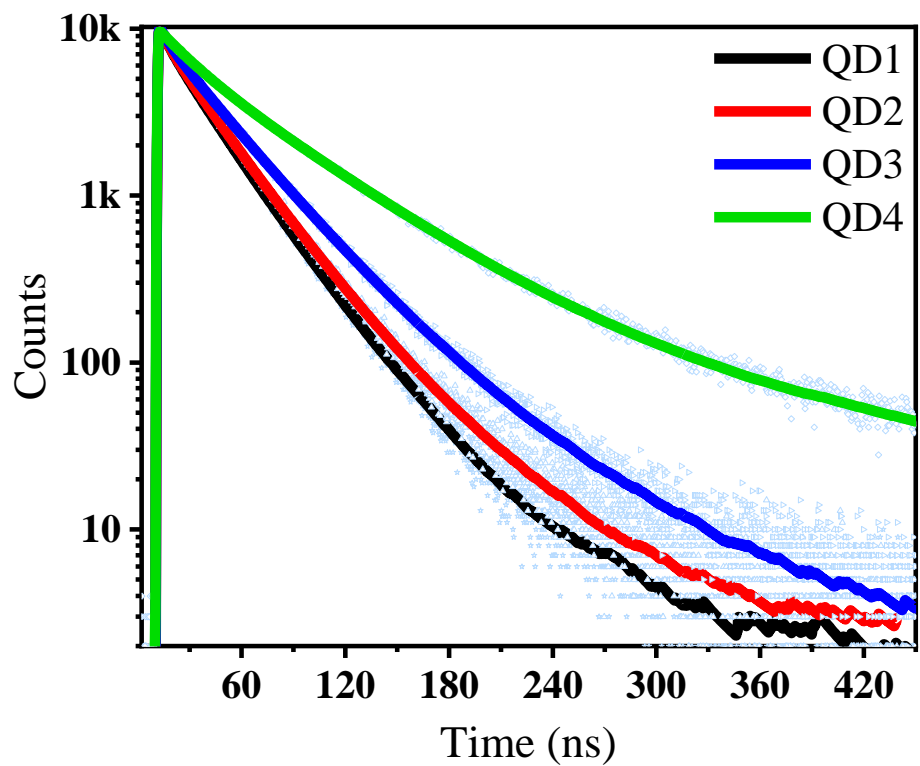

**Fig. S8:** Fluorescence lifetime of QDs excited with 454 nm with emission collected at the $\lambda_{em}^{max}$. The $\tau_{avg}$ was found to be 24.85 ns for QD1, 26.37 for QD2, 31.43 for QD3, and 52.53 ns for QD4.